# 1.5 bit-per-stage 8-bit Pipelined CMOS A/D Converter for Neuromophic Vision Processor

Yilei Li, Li Du

09212020027@fudan.edu.cn

*Abstract*- Neuromorphic vision processor is an electronic implementation of vision algorithm processor on semiconductor. To image the world, a low-power CMOS image sensor array is required in the vision processor. The image sensor array is typically formed through photo diodes and analog to digital converter (ADC). To achieve low power acquisition, a low-power mid-resolution ADC is necessary. In this paper, a 1.8V, 8-bit, 166MS/s pipelined ADC was proposed in a 0.18 um CMOS technology. The ADC used operational amplifier sharing architecture to reduce power consumption and achieved maximum DNL of 0.24 LSB, maximum INL of 0.35 LSB, at a power consumption of 38.9mW. When input frequency is 10.4MHz, it achieved an SNDR 45.9dB, SFDR 50dB, and an ENOB of 7.33 bit.

## I. INTRODUCTION

CMOS technology has widely been used in various electronic systems, such as human and machine interface, memory storage, communications, display, control, navigation and camera system [1-13]. Among those applications, camera based computer vision application have gained widely interest from both industry and academy due to its close relation with neural network. Recently, more and more computer vision algorithms have been implemented using hardware acceleration, resulting a much faster processing speed and low power property. To achieve an efficient vision system, CMOS front end sensor array is also required to be low power, thus to request an efficient ADC as the front end sensing block to bridge the analog signal and digital signal processing world. ADC is a device that converts a continuous physical voltage to a digital number that represents the quantity's amplitude. Pipelined structure, as one of the typical architectures has been widely implemented in the ADC design. A low power, middle-resolution (7~10 bit), middle speed (20MHz-200MHz) pipelined ADC is an important block in modern applications of telecommunication, consumer electronics, and medical electronics. Various power reduction techniques have been developed for pipelined ADCs, such as gain calibration for the sample and hold amplifier, flash ADC-based MSB quantization, and removing the front-end sample/hold amplifiers [14]. However, these techniques are difficult to apply on the middle resolution and middle speed pipelined ADCs due to their speed and resolution requirement.

A pipelined ADC working at this region generally consumes large power due to stringent requirements on capacitance mismatch and amplifier gain ([15]). Capacitance mismatch and low amplifier gain leads to both linear and nonlinear error in the multi-bit digital-to-analog converter (MDAC), which causes pipelined nonlinearity. The typical way to reduce this non-linearity is to use large capacitance in the MDAC, which usually causes the amplifier to consume large power to drive it. Besides, this solution still requires the amplifier to have enough gain to reduce the gain error.

A group of calibration techniques have been developed to compensate the most significant MDAC gain error. In [16-17], a reference ADC was used to calibrate a single nonlinear MDAC by estimating its 3rd-order harmonic term. Complicated adaptation algorithms, such as least-mean-square (LMS), however, are needed for the estimation of MDAC parameters. In [18], a digital processor with 8.4 K gates was implemented in order to use the adaption algorithms and it consumes a large area. The LMS loop also leads to a trade-off between step size and convergence speed. Besides, the adaption-based calibration techniques are not scalable as technology improves.

In this paper, an operational amplifier-sharing architecture has been implemented in the pipelined ADC. This amplifier-sharing architecture has halves the necessary number of the amplifiers without sacrificing system performance, and the power cost has been reduced too. A two-phase non-overlapping clock has been created to manage the amplifier's connection in different phase. To further increase the bandwidth, an improved switch-capacity-based common-mode feedback circuit has been embedded in the ADC.

The paper is organized in five sections. The system-level architecture of the pipelined ADC is introduced in Section II. Section III describes each building block design and performance. Simulation results are given in Section IV. Finally, Conclusions are drawn in Section V.

## II. SHARING AMPLIFIER PIPELINE ADC ARCHITECTURE

### A. System Level Architecture

In order to minimize the power consumption and increase the speed as much as possible, an operational transconductance amplifier (OTA) sharing architecture for two series stages have been implemented based on [19]. The fundamental benefit is the fact that for pipelined ADC, the OTA is only working in residue amplifier mode. During sampling mode, the OTA is relaxed, as Fig. 1 shows. Meanwhile, we notice that each following stage is half of a cycle delayed compared to the current stage. So when the first stage is in sampling mode, the second one is in residue amplifying mode, and versa vice. For this reason, the number of OTAs can be reduced by using one OTA for both stages, the single-ended architecture is shown Fig 2.

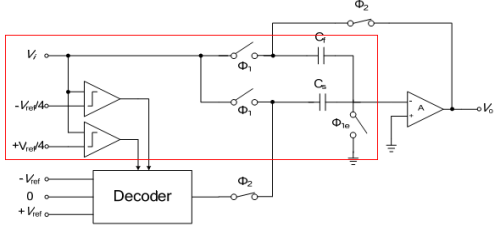

Fig.1 sampling mode of a pipeline stage

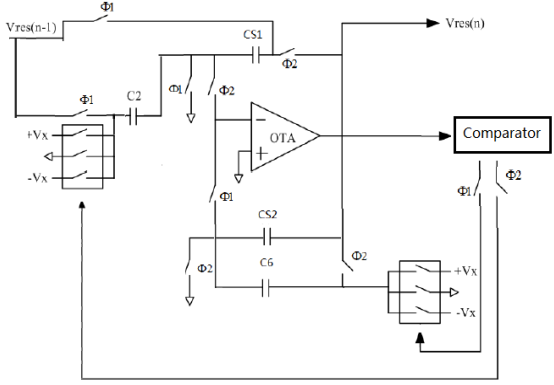

Fig.2 schematic of the sharing OTA

Based on this OTA sharing architecture, the whole system architecture can be drawn as Fig.3. The ADC consists of a sample and hold amplifier in the front, and six 1.5bit stage with OTA sharing architecture in the middle. At last, a 2bit-flash stage is implemented to quantify the remaining 2 bits.

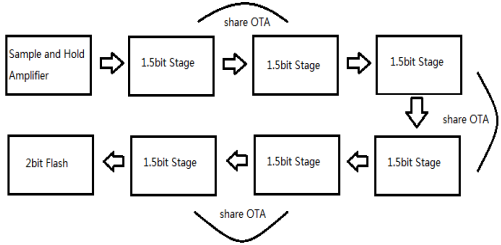

Fig.3 whole ADC architecture

### B. Op-Amp Gain and Speed Calculation

In pipeline ADC, the sample-mode and hold-mode circuit in each stage can be drawn as Fig.4. In order to get integral nonlinearity (INL) <1LSB, we need to make sure that the settling error of each stage is less than 1/2 least significant bit (LSB) [21]. The settling voltage of each stage can be calculated as:

$$V_{out} = V_{static}\left(1 - e^{-\frac{t}{\tau}}\right) \quad [1]$$

where

$$V_{static} = \frac{A_0}{1+\beta A_0}V_{in} \quad [2]$$

$$\tau = \frac{1}{\omega_c} = \frac{C_{Load}}{\beta g m} = \frac{1}{2\pi\beta GBW} \quad [3]$$

As the equation [2] [3] shows, the final setting voltage has both static setting error and the dynamic error due to the limit speed and gain of the OTA.

If we assign the 1/2LSB equally to both static and dynamic error, we derive the following equations:

$$E_{gain} = \frac{1}{\beta A_0} < 1/4\text{LSB} \quad [4]$$

$$E_{dyna} = e^{-t \times 2\pi\beta GBW} < 1/4\text{LSB} \quad [5]$$

Based on the sampling rate of the ADC (166MHz), we can calculate the required OTA's DC gain should be more than 67dB while its gain-bandwidth-product(GBW) needs to be larger than 950MHz.

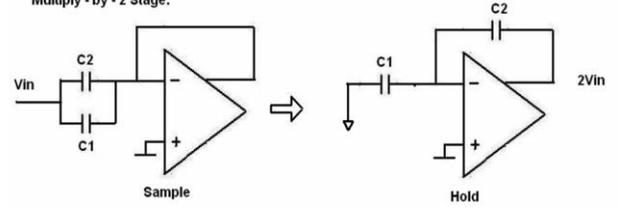

Fig.4 sample-mode and hold-mode circuit of a stage

### C. Sample and Hold Circuit

Although some paper [20] mentions about designing pipelined ADC without sample and hold amplifier (SHA) in front, it is not suitable for a high performance ADC design due to the different delay between two paths.

To address this issue, the switch capacitor based sample and hold circuit is implemented as Fig.5. In ck1, the circuit samples the signal voltage and stored into the sampling capacitor. In ck2, the OTA's input and output switches are turned on and the OTA forms a feedback loop. Thus the sampling capacitor's negative side voltage is set to 0 and the OTA generates the sampled voltage at its output.

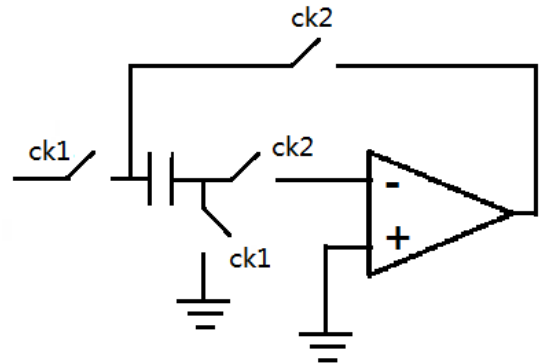

Fig.5 sample and hold circuit

### III. BUILDING BLOCK DESIGN AND SIMULATION

#### A. OTA Design

A typical telescopic amplifier with gain boosting structure has been implemented in the OTA design. The main amplifier's gain is boosted by incorporating two boosting amplifiers with relative low current consumption. One of the boosting amplifier has been implemented as folded cascade structure in order to support large output swing voltage at the OTA's output. The simulated DC gain can go up to 85dB and the GBW is up to 2.5GHz as shown in Fig.7.

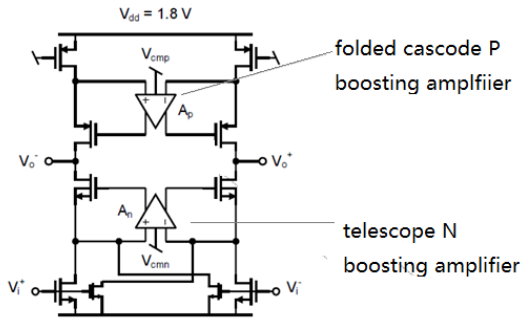

Fig.6 amplifier structure

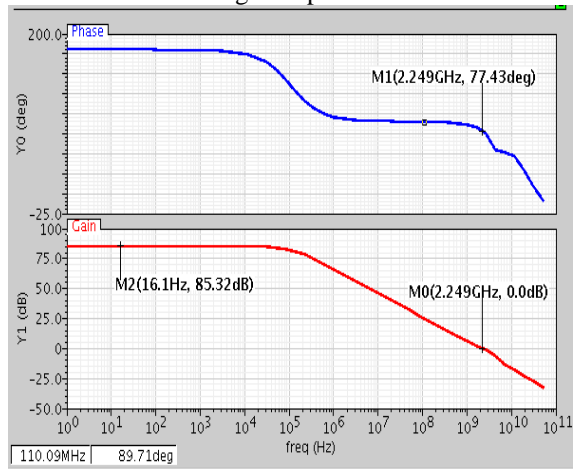

Fig.7 AC performance of the amplifier

### B. Improved Switch-Capacity-Based Common Mode Feedback Circuit

In order to make sure the amplifier works correctly, its output common voltage needs to be set correctly. Because the amplifier works in both clock cycles now, a traditional switch capacity common mode feedback (CMFB) circuit is not feasible, as shown Fig 7.1. So an improved CMFB circuit has been proposed, as shown in Fig 7.2. The benefit of this circuit is that at whatever point in clk1 or clk2 cycle, the amplifier's common mode voltage is set properly.

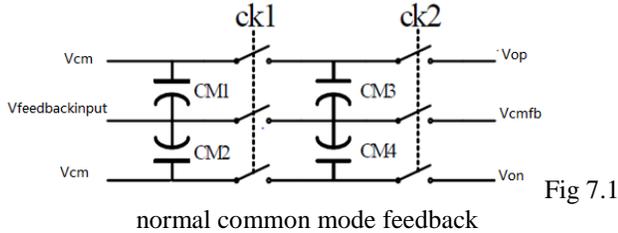

Fig 7.1 normal common mode feedback

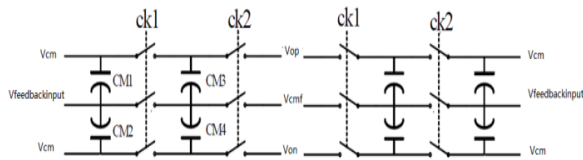

Fig 7.2 improved common mode feedback

### C. Comparator Design

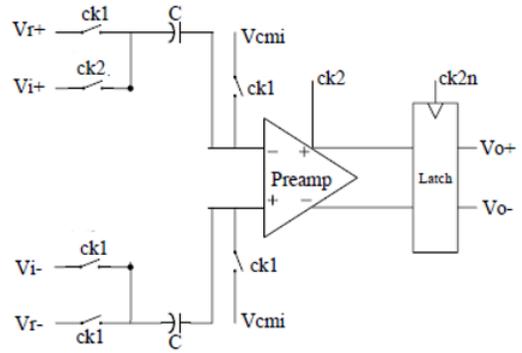

Fig.8 comparator circuit

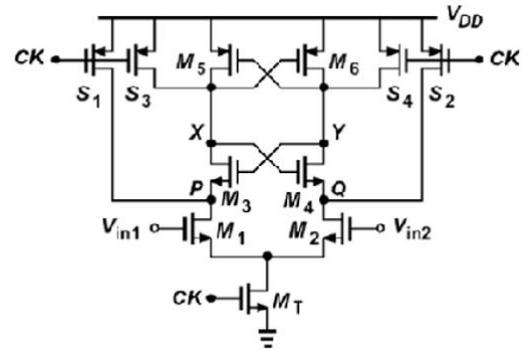

Fig. 9 pre-amplifier circuit

Fig.8 and Fig. 9 show the switch-capacity-based comparator's circuit and its pre-amplifier architecture. Because the charge stored in C is constant, we can derive the pre-amplifier's input differential voltage as follows:

$$V_{diff} = (V_{r+} - V_{r-}) - (V_{i+} - V_{i-}) \quad [7]$$

The pre-amplifier is implemented as a strong-arm architecture. S1 and S2 have been placed at input transistors' output to reset the amplifier during the regeneration mode. M5, M6, M3, M4 generate two positive feedback circuits used to accelerate the comparator speed.

### D. Clock Arrangement for Sharing Amplifier Structure.

Due to the OTA works in both clock cycles, a pair of non-overlapping clocks is needed. Besides this, an additional clock period between the two working periods has been generated for reset, which is used for resetting the OTA (connecting its two differential outputs to common mode voltage). This can diminish the residue charge on the amplifier output node, which may affect the result of the next cycle amplification. The clock phase is shown in Fig. 10.

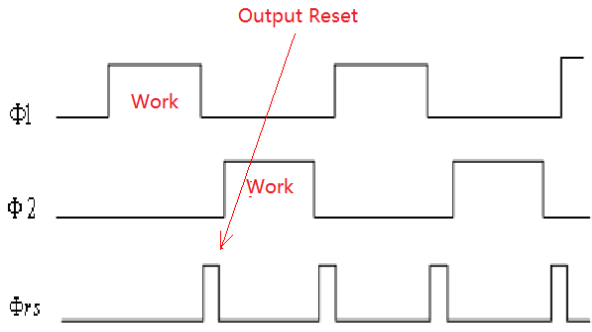

Fig.10 clock arrangement

As shown in Fig.10, the OTA will work during ∅1 and ∅2 period and reset in ∅rs period. The ∅rs is generated through a NOR calculation of the ∅1 and ∅2. To compensate the delay of the NOR gate, an inverter has been placed after ∅1 and ∅2 to make these three clocks synchronous.

## IV. ADC SIMULATION RESULT

### A. Stage Setup Simulation

In order to test the ADC's sample and hold performance, an input pulse from -600mV to 600mV has been injected into the ADC. In this case, the amplifier output needs to change from -600mV to 600mV in one cycle. The simulation result is listed as Tab.2.

| Stage | Ideal/mV | Simulation/mV | Error% |
|---|---|---|---|
| Vin | 600 | 600 | 0 |
| SHA | 600 | 599.7 | 0.05% |
| Stage1 | 599.7 | 599.2 | 0.08% |
| Stage2 | 599.2 | 598.5 | 0.11% |
| Stage3 | 598.5 | 596.3 | 0.36% |
| Stage4 | 596.3 | 593.9 | 0.4% |
| Stage5 | 593.9 | 587.4 | 1.1% |
| Stage6 | 587.4 | 575.6 | 2% |

Tab.2 Stage setup time simulation

As shown in the Tab.2, the input signal has been setup accurately (error<1%) till 5th stage. The 5th stage and 6th stage have relative large setup error due to the accumulation of the errors from previous stages. However, at that stage, the requirement of the accuracy is relative low because most of the MSB have already been quantified.

### B. Linearity Simulation

To measure the INL and differential nonlinearity (DNL) of the ADC, a very slow ramp signal has been injected to the ADC input. The measured INL and DNL is shown Fig.11. Based on the curve, maximum INL is 0.35LSB happened at code 224 and maximum DNL is 0.24LSB happened at code 96.

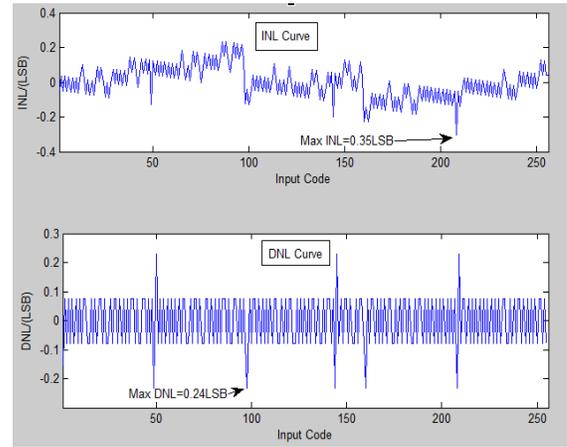

Fig.11 ADC INL DNL measurement

### C. Top Level Performance Simulation

To verify the ADC's top level performance, the ADC was running at a sampling frequency of 166.6MHz and a 10.417MHz sinusoidal signal was injected into the ADC input. The measured output spectrum is shown in Fig.12. Based on this, the calculated signal-to-noise and distortion ratio (SNDR) is 45.9dB，Spurious-Free Dynamic Range (SFDR) is 50dB, and Effective number of bits (ENOB) is equal to 7.33 bit.

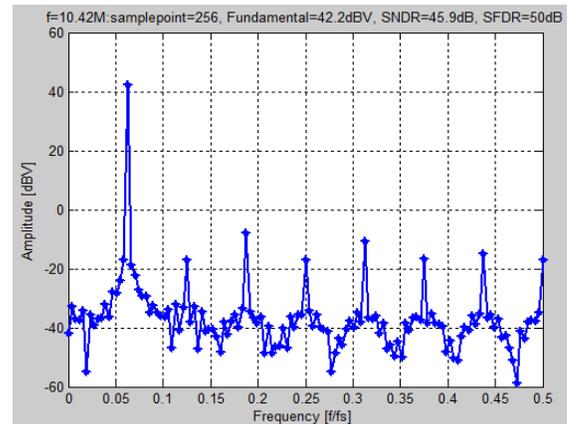

Fig.12 Spectrum of the ADC output

## V. CONCLUSION

In this paper, an 8bit 166MS/s 38.9mW pipeline ADC with OTA sharing architecture is proposed. In order to let the OTA work in both cycles, an improved CMFB circuit is implemented and generates another clock $\emptyset_{reset}$ to diminish residue charge on the output of the amplifier. Simulation results show that the ADC has INL of 0.35LSB and DNL of 0.24LSB. When the input signal is 10.42MHz and sampling frequency is 166MS/s, it achieves an SNDR of 45.9dB, SFDR of 50dB and an ENOB of 7.33bit.